\documentclass[article]{aa}
\usepackage{natbib}
\usepackage{longtable,lscape}
\usepackage{amsmath}
\usepackage{natbib}
\bibpunct{(}{)}{;}{a}{}{,}
\usepackage{graphicx}
\usepackage{multirow}

\usepackage{graphicx}
\usepackage{natbib}
\usepackage[colorlinks=true,citecolor=blue]{hyperref}
\usepackage{color}
\usepackage{xspace}
\usepackage{dcolumn}
\usepackage{longtable}
\usepackage{lscape}
\usepackage{soul}
\usepackage{wasysym}
\usepackage{rotating}
\usepackage{afterpage}
\usepackage{mathtools}
\usepackage{blindtext}

\newcommand{\MJup}{M$_{\mathrm{Jup}}$\xspace}
\newcommand{\RJup}{R$_{\mathrm{Jup}}$\xspace}
\newcommand{\RSun}{R$_{\odot}$\xspace}

\newcommand{\mic}{$\mu$m\xspace}

\usepackage[varg]{txfonts}
\begin{document}
%
    \title{Characterizing HR\,3549\,B using SPHERE} 


    \author{D. Mesa\inst{1}, A. Vigan\inst{2}, V. D'Orazi\inst{1,3,4}, C. Ginski\inst{5}, S. Desidera\inst{1}, M. Bonnefoy\inst{6,7}, 
     R. Gratton\inst{1}, M. Langlois\inst{8,2}, F.Marzari\inst{9,1}, S. Messina\inst{10},
     J. Antichi\inst{11,1}, B. Biller\inst{12,13}, M. Bonavita\inst{12,1}, E. Cascone\inst{14}, G. Chauvin\inst{6,7}, R.U. Claudi\inst{1}, 
     I.Curtis\inst{15}, D. Fantinel\inst{1}, M. Feldt\inst{13}, A. Garufi\inst{16}, R. Galicher\inst{17}, Th. Henning\inst{13}, 
     S. Incorvaia\inst{18}, A.-M. Lagrange\inst{6,7}, M. Millward\inst{19}, C. Perrot\inst{17}, 
     B. Salasnich\inst{1}, S. Scuderi\inst{10}, E. Sissa\inst{1}, Z. Wahhaj\inst{20,2}, A. Zurlo\inst{21,22,2}}

   \institute{\inst{1}INAF-Osservatorio Astronomico di Padova, Vicolo dell'Osservatorio 5, Padova, Italy, 35122-I \\
   \inst{2}Aix Marseille Universit\'e, CNRS, LAM - Laboratoire d'Astrophysique de Marseille, UMR 7326, 13388, Marseille, France\\
   \inst{3}Department of Physics and Astronomy, Macquarie University, Sydney, NSW 2109, Australia \\
   \inst{4}Monash Centre for Astrophysics, School of Physics and Astronomy, Monash University, Melbourne, VIC 3800, Australia \\
   \inst{5}Leiden Observatory, Leiden University, PO Box 9513, NL-2300 RA Leiden, the Netherlands \\
   \inst{6}Universit\`e Grenoble Alpes, IPAG, F-38000 Grenoble, France \\  
   \inst{7}CNRS, IPAG, F-38000 Grenoble, France  \\
   \inst{8}CRAL, UMR 5574, CNRS, Universit\'e Lyon 1, 9 avenue Charles Andr\'e, 69561 Saint Genis Laval Cedex, France \\
   \inst{9}Department of Physics, University of Padova, I-35131 Padova, Italy \\
   \inst{10}INAF-Osservatorio Astrofisico di Catania, Via S. Sofia 78, 95123, Catania, Italy \\
   \inst{11}INAF-Osservatorio Astrofisico di Arcetri – L.go E. Fermi 5, 50125 Firenze, Italy \\
   \inst{12}Institute for Astronomy, University of Edinburgh, Blackford Hill View, Edinburgh EH9 3HJ, UK \\
   \inst{13}Max-Planck-Institut f\"ur Astronomie, K\"onigstuhl 17, 69117 Heidelberg, Germany \\
   \inst{14}INAF-Astrophysical Observatory of Capodimonte, Salita Moiariello 16, 80131 Napoli, Italy \\
   \inst{15}YSVP Observatory, 2 Yandra Street Vale Park, South Australia 5081, Australia \\
   \inst{16}Institute for Astronomy, ETH Zurich, Wolfgang-Pauli-Strasse 27, CH-8093 Zurich, Switzerland \\ 
   \inst{17}LESIA, Observatoire de Paris, PSL Research Univ., CNRS, Univ. Paris Diderot, Sorbonne Paris Cit\'e, UPMC Paris 6, Sorbonne Univ., 5 place Jules Janssen, 92195 Meudon CEDEX, France\\ 
   \inst{18}INAF-Istituto di Astrofisica Spaziale e Fisica Cosmica di Milano, Via E. Bassini 15, 20133 Milano, Italy \\
   \inst{19}York Creek Observatory, Georgetown, Tasmania, Australia \\
   \inst{20}European Southern Observatory, Alonso de Cordova 3107, Vitacura, Santiago, Chile \\
   \inst{21}N\'ucleo de Astronom\'ia, Facultad de Ingenier\'ia, Universidad Diego Portales, Av. Ejercito 441, Santiago, Chile \\
   \inst{22}Universidad de Chile, Camino el Observatorio, 1515 Santiago, Chile \\
}

   \date{Received  / accepted }

\abstract
   {}
   {In this work, we characterize the low mass companion of the A0 field star HR\,3549.}
   {We observed HR\,3549AB in imaging mode with the
the NIR branch (IFS and IRDIS) of SPHERE@VLT, with IFS in $YJ$ mode and IRDIS in the H band. 
We also acquired a medium resolution spectrum with the IRDIS long slit
spectroscopy mode.  The data were reduced using the dedicated SPHERE
GTO pipeline, purposely designed for this instrument.  We employed
algorithms such as PCA and TLOCI to reduce the speckle noise.}
    {The companion was clearly visible both with IRDIS and IFS. We
      obtained photometry in four different bands as well as
      the astrometric position for the companion.  Based on our
      astrometry, we confirm that it is a bound object and put constraints on its
     orbit.   Although several uncertainties are still present, we
     estimate an age of $\sim$100-150 Myr for this system, yielding a 
 most probable mass for the companion of 40-50\MJup and $T_{eff}\sim2300-2400$\,K. Comparing 
     with template spectra points to a spectral type between M9 and L0 for the companion, 
     commensurate with its position on the color-magnitude diagram.}
{}


\titlerunning{Characterizing HR\,3549\,B with SPHERE}
\authorrunning{Mesa et al.}
   \maketitle
%

\section{Introduction}
\label{intro}
In recent years, a handful of giant planets and brown dwarfs have been
discovered around young stars (with ages less than few hundreds of 
Myr) through the direct imaging technique (see e.g. \citealt{2005A&A...438L..29C}, 
\citealt{2005A&A...438L..25C}, \citealt{2008Sci...322.1348M}, \citealt{2010Natur.468.1080M}, 
\citealt{2010Sci...329...57L}, \citealt{2010ApJ...720L..82B}, \citealt{2013ApJ...763L..32C}, 
\citealt{2013ApJ...772L..15R}, \citealt{2014ApJ...780L...4B}).
However, it is difficult to characterize these objects, as they
generally lack precision multiwavelength photometry and spectrometry.  Thus, fundamental properties such as e.g. mass, radius, $T_{eff}$, $\log{g}$ and spectral type are often
poorly constrained for these objects. \par
The low-mass companion to the main sequence A0 star HR\,3549 (HIP\,43620;
HD\,76346) is such an object. 
Discovered by \citet{Mawet2015} in $L'$-band observations with NACO at
VLT, the detected companion was at a separation
of $\sim$0.9 arcsec and at a position angle of $\sim$$157^{\circ}$ in
the discovery epoch.
HR\,3549 has a parallax of 10.82$\pm$0.27 mas \citep{vanleeuwen2007} implying a distance of $\sim$92.5 pc.  
However, while the distance of the system is well-constrained, the
host star has an estimated age between 50 and 400 Myr, leading to a wide range both for the mass 
estimation (between 15 and 90 \MJup) and for the effective temperature
of the companion (between 1900 and 2700 K). 
Thus, it was not possible to precisely infer the fundamental
properties of the companion and it was generically identified as an L-type object. 
The parent star hosts a dust disk as well, based on a measured WISE infrared
excess at 22~\mic (W1-W4=0.56$\pm$0.06 mag)
\citep{2012yCat.2281....0C,Mawet2015}.  No excess was found at
12~\mic, implying a temperature for the dust disk of 153\,K and an
outer disk radius of $\sim$17 AU from the host star, well inside the
companion position. \par   
                       
In the last few years, a new generation of direct-imaging instruments
have come online.  These instruments provide both precise multiband photometry and low and medium resolution
spectroscopic capabilities, enabling a better characterization of
low-mass companions to young stars close to the Sun. SPHERE at the VLT is a member of this cohort of
instruments \citep{2008SPIE.7014E..18B} and started operations at the beginning of 2015.
It is composed of three scientific modules operating both in the NIR with IRDIS \citep{Do08} and IFS \citep{Cl08}, and
in the visible with ZIMPOL \citep{Th08}. It is equipped with the SAXO extreme adaptive optics system 
\citep{Fusco:06, 2014SPIE.9148E..0OP, doi:10.1117/12.2056352}, with a 41$\times$41 actuators wavefront control, 
pupil stabilization, differential tip tilt control and employs stress polished toric mirrors for beam 
transportation \citep{2012A&A...538A.139H}.  In the main IRDIFS
imaging mode, low resolution (R=50) spectra are obtained with the IFS
between 0.95 and 1.35 \mic while IRDIS is simultaneously used in dual-band imaging mode (DBI; \citealt{Vig10}) with the H23
filter pair (wavelength $H2$=1.587 \mic; $H3$=1.667 \mic).
A lower resolution (R=30) spectrum but with a wider spectral coverage can be obtained when
SPHERE is operating in the IRDIFS\_EXT mode.  In this mode, R=30 spectra
are obtained with the IFS in the YH band between 0.95 and 1.65 \mic
while IRDIS is simultaneously used in dual-band imaging mode 
with the K12 filter pair (wavelength $K1$=2.110 \mic; $K2$=2.251
\mic). A more complete characterization
of the companions can be carried out with IRDIS in the long slit spectroscopy (LSS) mode that supplies 
a medium resolution spectrum (MRS - R=350).
In the past year, SPHERE has demonstrated its capability to characterize
substellar companions in e.g. \citet{2016A&A...587A..56M}, \citet{2016A&A...587A..55V}, \citet{2016A&A...587A..58B}, 
\citet{2016A&A...587A..57Z} and \citet{2015ESS.....320305B}. \par 
We observed HR\,3549 with the NIR branch instruments of SPHERE in December 2015. In this paper we report our results on these
observations. In Section~\ref{s:obs} we describe observations and data reduction, 
in Section~\ref{s:res} we illustrate the results that are then discussed in Section~\ref{s:dis}. Finally, in 
Section~\ref{conclusion} we provide our conclusions.

\section{Observations and data reduction}
\label{s:obs}

HR\,3549 was observed on December 19th 2015 with SPHERE operating in
the IRDIFS mode. For both IRDIS and IFS 
the dataset was composed of 16 datacubes, each of them with 4 individual frames of 64 s exposure time. 
The IRDIS observations used a 4x4 pixels dithering pattern while no dithering
was used for the IFS observation. To enable angular differential imaging \citep[ADI;][]{2006ApJ...641..556M} 
technique, the field of view (FOV) was allowed to rotate during the
observations. To maximize the rotation, we observed over the meridian
passage of the star, for a total FOV rotation of $\sim$$32.5^{\circ}$. For both instruments, frames with the point spread function (PSF) off-centered
with respect to the coronagraph and frames with four satellite spots symmetric with respect to the central 
star were also taken before and after the coronagraphic observations to allow proper flux calibration and 
centering of the frames with respect to the star. The use of satellite spots to define
the center of an image was first proposed by \citet{2006ApJ...647..620S} and \citet{2006ApJ...647..612M} and its
use in SPHERE is explained in \citet{2013aoel.confE..63L} and in \citet{mesa2015}. 
To avoid saturation, the off-centered frames were observed using a neutral density filter.\par
The same target was observed again in the night of December 27th 2015 with 
IRDIS operating in long slit spectroscopy \citep{vigan2008} mode. In this case the dataset was 
composed of 23 datacubes, each of them composed of 5 frames with an exposure time of 32 s. The sequence also 
included the acquisition of an off-axis reference PSF by moving the star off of the coronagraph. To avoid saturation, 
this off-axis PSF was acquired using a neutral density filter \citep[see e.g.][]{vigan2015}. We
used IRDIS-LSS in MRS corresponding to R=350. 

\subsection{IRDIFS data reduction}
\label{s:irdifsdatared}

Data reduction for the IFS data was performed following the procedure
described in \citet{mesa2015} and in \citet{zurlo2014}.
We applied the appropriate calibrations (dark, flat, spectral positions, wavelength calibration and instrument flat)
to create a calibrated datacube composed of 39 images of different wavelengths for each frame obtained during the
observations. The calibrated datacubes were then registered and combined using the principal 
components analysis \citep[PCA; e.g. ][]{2012ApJ...755L..28S} algorithm to implement
both ADI and spectral differential imaging \citep[SDI,][]{1999PASP..111..587R} to remove the speckle noise.
For IRDIS, after the application of the appropriate calibrations (dark, flat and centering), the speckle subtraction was 
performed using both the PCA and the TLOCI \citep{2014SPIE.9148E..0UM} algorithms.
For both IFS and IRDIS the data reduction was partly performed using the pipeline of the SPHERE data center
hosted at OSUG/IPAG in Grenoble. \par 
Given the contrast of the order of $10^{-4}$ and the separation larger than 0.8 arcsec, the companion is visible
even in a simple de-rotated and stacked image.  As will be discussed in the next Sections, 
this helps calibrate and account for companion self-subtraction
produced by the PCA and TLOCI algorithms.

\subsection{IRDIS/LSS data reduction}

The LSS data was analyzed using the SILSS pipeline
\citep{2016ascl.soft03001V}, which has been developed specifically to
analyse IRDIS LSS data. The pipeline combines the standard ESO
pipeline with custom IDL routines to process the raw data into 
a final extracted spectrum for the companion. After creating the static
calibrations (background, flat field, wavelength calibration), the
pipeline calibrates the science data and corrects for the bad
pixels. It also corrects for a known issue of the MRS data, which
produces a variation of the PSF position with wavelength because of a
slight tilt ($\sim$1 degree) of the grism in its mount. To correct for
this effect, the pipeline measures the position of the off-axis PSF in
the science data as a function of wavelength, and shifts the data in
each spectral channel by the amount necessary to compensate for the
chromatic shift. All individual frames are calibrated independently
for the two IRDIS fields.

After this calibration and correction step, the speckles are
subtracted in the data following an approach based on the spectral
differential imaging technique described in \citet{vigan2008} and
\citet{vigan2012a}. The method has now been optimized to provide
a better subtraction of the speckles: instead of constructing a simple
reference of the stellar signal as the mean (or median) of all the
spatial channels where the signal of the planet is not present, we use
all spatial channels where there is no signal from the companion as
reference, and subtract a linear combination of these references to
each of the channels where the companion is present. To best reproduce
and subtract the speckles, the coefficients for the linear combination
are optimized on the areas where the signal of the companion is
absent. In practice, this approach is similar to the locally-optimized
combination of images \citep[LOCI;][]{lafreniere2007} applied to
LSS. This analysis is performed on all frames independently, and then
the speckle-subtracted frames are median-combined. Since the
wavelength calibration is slightly different for both IRDIS fields, we
obtain a final frame for each of the two fields.

The extraction of the spectrum of the companion in both IRDIS fields
is performed using a 1~$\lambda/D$-wide aperture in each spectral
channel. The exact separation of the companion within the slit is
known, but we optimized the position of the aperture so as to maximize
the final integrated signal. The noise is measured by integrating the
residual signal at a symmetric position with respect to the star,
i.e. at a location where the speckles have been subtracted but where
there is no companion signal. The spectrum of the off-axis reference
PSF is extracted using an aperture of the same width.  For the
reference PSF, the effect of the neutral density filter is compensated
in each spectral channel using a dedicated
tool\footnote{\url{http://people.lam.fr/vigan.arthur/tools.html}}. The
spectrum of the companion calibrated in contrast is then obtained by
dividing its extracted spectrum by that of the off-axis reference
PSF. Finally, the spectra obtained for each of the two IRDIS fields
are interpolated on the same wavelength grid and averaged to increase
the signal-to-noise ratio of the final spectrum.

\subsection{Non SPHERE observations}
\label{s:nosphere}

We obtained photometric observations of HR 3549 over six nights
between 3-26 March 2016 in order to measure its rotation period.
We observed for four nights at the YSVP Observatory (Vale Park, South Australia, $-$34$^{\circ}$53$^{\prime}$04$^{\prime\prime}$; 
138$^{\circ}$37$^{\prime}$51$^{\prime\prime}$E; 44\,m a.s.l.) using a 23-cm Schmidt-Cassegrain telescope. We collected 1600 
frames in the Johnson R-band filter, in defocused mode and with a per-frame exposure time of 1.3 s exposure.
On each night, observations were collected for up to 8 consecutive
hours, achieving an average photometric precision of $\sigma_R$ = 0.005\,mag.
We observed for two nights at the York Creek Observatory (YKO, Launceston, Tasmania, Australia, 
$-$41$^{\circ}$06$^{\prime}$06$^{\prime\prime}$; 146$^{\circ}$50$^{\prime}$33$^{\prime\prime}$E; 28\,m a.s.l) using  a 
25-cm Takahashi Mewlon reflector. We collected 32 frames in the Johnson R-band filter, using 1 s exposures. 
Observations were collected for up to 5 consecutive hours, achieving
an average photometric precision of $\sigma_R$ = 0.005\,mag.\par
Bias subtraction,  flat field correction, and aperture magnitude extraction were done using IRAF routines. 
We built an ensemble comparison star using four nearby stars that were
found to have constant flux;
Differential R-band magnitudes of HR\,3549 were then obtained relative
to this comparison star.


\section{Results}
\label{s:res}

\subsection{IRDIFS}
\label{s:irdifs}

\begin{center}
\begin{figure*}[!htp]
\centering
\includegraphics[width=0.45\textwidth]{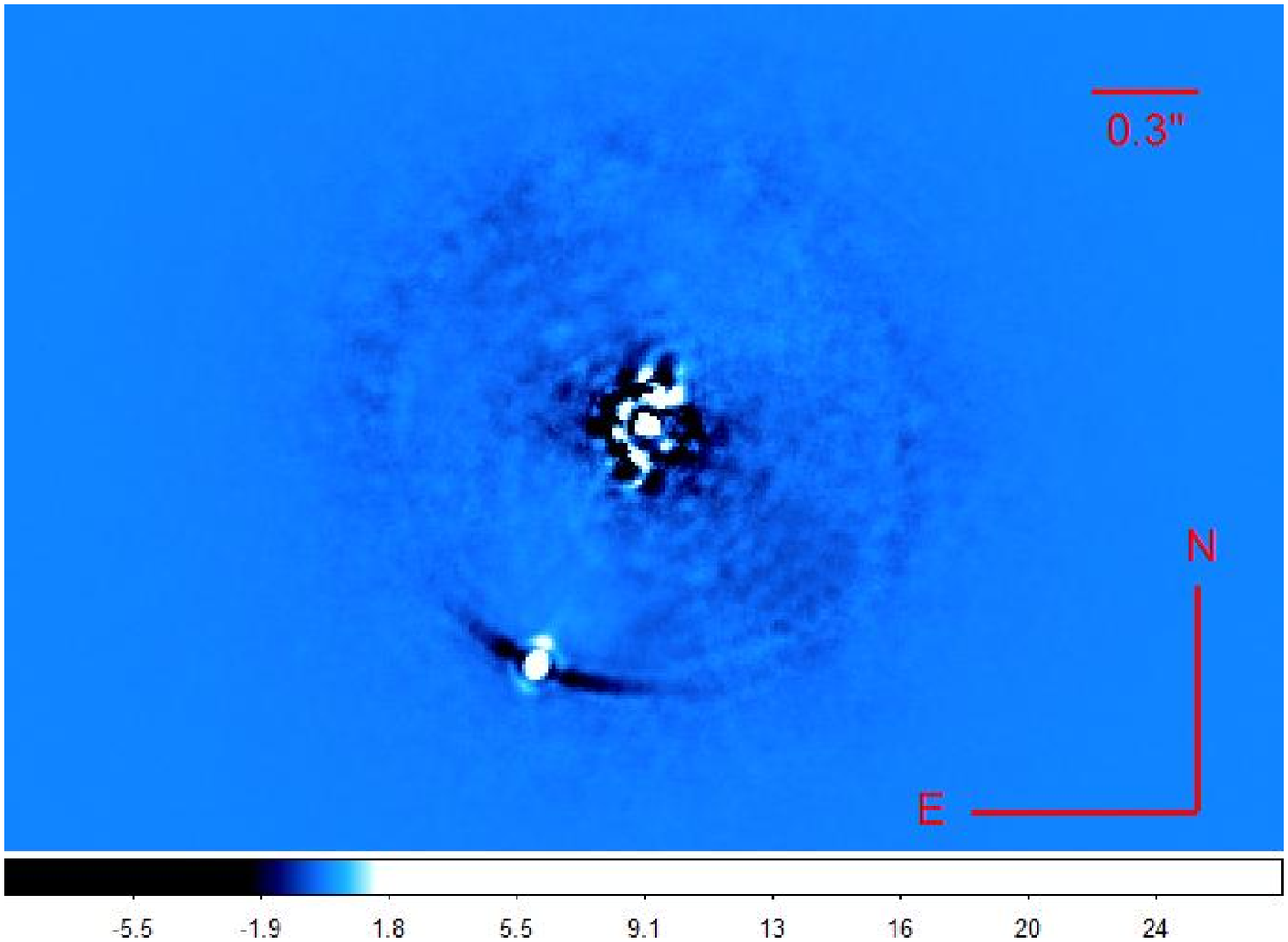}
\includegraphics[width=0.45\textwidth]{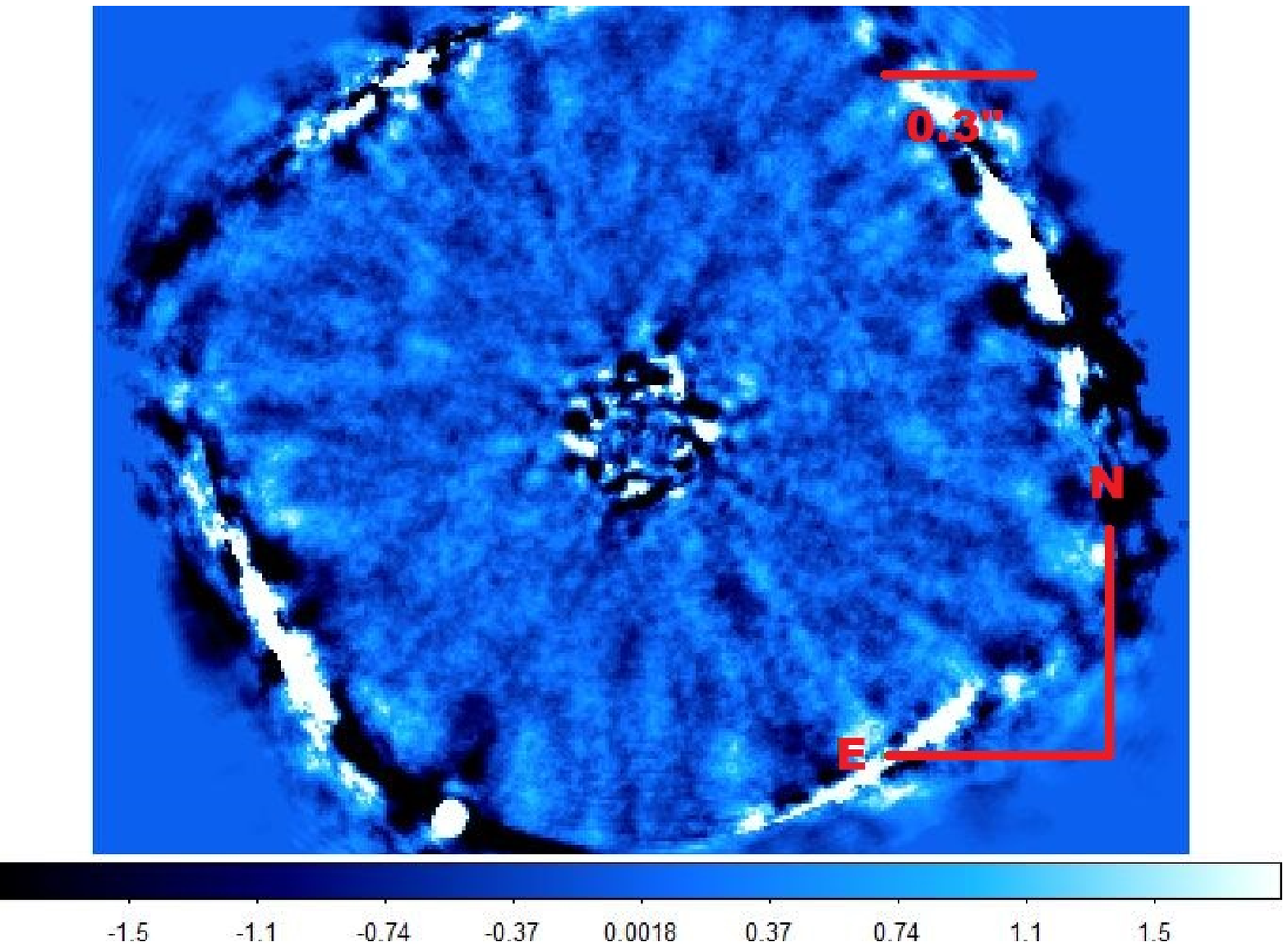}
\caption{{\it Left:} Final image obtained with IRDIS.{\it Right:} Final image obtained with IFS.}
\label{f:finalim}
\end{figure*}
\end{center} 

In Figure~\ref{f:finalim} we display the final image obtained with IRDIS (left) and with IFS (right). While in both
cases the companion is clearly visible, in the IFS case it is just at
the edge of the instrument field of view (FOV), 
which introduces some difficulties in extracting photometry for this object.
The companion position was measured by inserting negative scaled
PSF images into the final image and shifting the simulated companion position until the
standard deviation was minimized in a small region around the companion.
This procedure was repeated in the final
images obtained with differing numbers of principal components in the
PCA analysis (as described in Section~\ref{s:irdifsdatared}) and the
error adopted on position was calculated as the standard deviation on
these measures. The dominant error source is different for separation
vs. position angle; the most important contribution to the the error
on the separation is the uncertainty on the centering of the host star
(assumed to be half of the pixel scale).
On the other hand, the main contribution to the error on the position
angle is the uncertainty on the position of the true north (TN),
calculated by observing an astrometric calibration field.
Astrometric measurements performed with IRDIS and IFS are listed in Table~\ref{t:astro}. 

\begin{table*}[!htp]
\caption{Astrometric position for HR\,3549\,B.}
\label{t:astro}
\centering
\begin{tabular}{c c c c c}
\hline
\hline
Instrument & $\Delta$$\alpha$ (arcsec) & $\Delta$dec (arcsec) &  Separation (arcsec) & Position angle \\
\hline   
   IFS   &  0.348$\pm$0.004    &  -0.776$\pm$0.004     &  0.850$\pm$0.006 &   155.8$\pm$0.5   \\
  IRDIS   & 0.344$\pm$0.007    &  -0.775$\pm$0.007     &  0.848$\pm$0.009 &   156.1$\pm$0.7   \\
\hline
\end{tabular}
\end{table*} 

\begin{figure}
\begin{center}
\centering
\includegraphics[width=8.0cm]{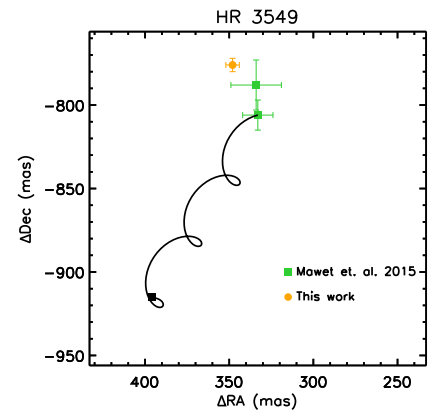}
\caption{Common proper motion analysis for HR\,3549\,B. The black solid line displays the motion of the companion
if it was a background object {\bf while the black square at the end of the line indicates the position that it would
have at the observation date}. The green points are taken from \citet{Mawet2015}, the orange
one is from our astrometric measurements. We only plot astrometric
values from the IFS observation, as our IFS observations provided a
higher astrometric precision than our IRDIS observations.}
\label{f:astro}
\end{center} 
\end{figure}

Exploiting our astrometric results we were able to extend the common proper motion analysis from \citet{Mawet2015}, further
confirming that HR\,3549\,B is a bound object. The result of this analysis is shown in Figure~\ref{f:astro}. We interpret 
the small changes in projected separation and position angle with respect to \citet{Mawet2015} as likely due to
to orbital motion. The possible orbital solutions compatible with the data are discussed in Section~\ref{s:orbit}.

In Table~\ref{t:photo} we report the photometry obtained in four different bands using
both IFS data for Y and J band and IRDIS data for H2 and H3 band taken
as described in Section~\ref{s:obs}. 
IFS photometry was obtained by median combining all spectral channels
between 0.95 and 1.15 $\mu$m for the Y band and between
1.15 and 1.35 $\mu$m for the J band. The error bars were calculated in
a similar way, but do not incorporate
the errors given by the uncertainty on the parallax given in Section~\ref{intro}. For this reason, an error of the 
order of 0.055 mag was to be added to the error bars listed in Table~\ref{t:photo}.

\begin{table}
\caption{Absolute magnitudes for HR\,3548\,B in four different spectral band from the IRDIFS data.}
\label{t:photo}
\centering
\begin{tabular}{c c c c}
\hline
\hline
  Y  & J  &  H2 & H3 \\
\hline   
 $11.98\pm0.16$    &  $11.06\pm0.06$     &  $10.24\pm0.07$  &  $10.14\pm0.02$    \\
\hline
\end{tabular}
\end{table} 

\begin{figure}
\begin{center}
\centering
\includegraphics[width=8.0cm]{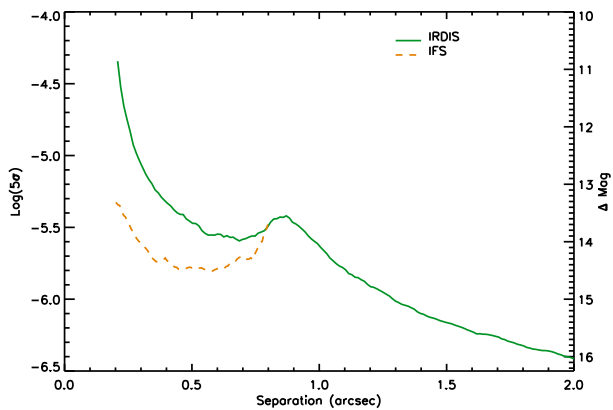}
\caption{Contrast plot for our HR\,3549 observations. IRDIS contrast
  is plotted with a green solid line while IFS contrast is plotted
  with an orange dashed line.  The contrast for both instruments was obtained using the PCA 
algorithm.}
\label{f:contrast}
\end{center} 
\end{figure}

The 5-$\sigma$ contrast curve derived from both IRDIS and IFS final images is
shown in Figure~\ref{f:contrast} where the green line is the contrast obtained
using IRDIS while the dashed orange line is the one obtained from IFS.   
At a separation of 0.5'', we obtained a contrast of the order of
$\sim1.6\times10^{-6}$ ($\Delta$J$\sim$14.4) with the IFS and a
contrast $\sim3.2\times10^{-6}$ ($\Delta$H$\sim$13.7) with IRDIS.
At separations $>$1.3'', IRDIS achieves contrasts better than $10^{-6}$.

As previously noted, the large separation and the relatively small contrast between the companion and the
host star allowed us to see the companion in the calibrated datacube
after only de-rotation and median-stacking.  The PCA algorithm used to
remove speckle noise does so at expense of removing some of the light
from the companion (aka self-subtraction).  Since the companion is
easily retrieved without PCA in this case, we can use the simple
de-rotated and median-stacked reduction to mitigate the effect of
self-subtraction from the PCA algorithm.  

Both LOCI and PCA algorithms build ideal PSFs and then subtract them
from the data image.  If these PSFs still contain light from the
companion, then some of the flux of the companion will be removed in
this subtraction.  We thus build our PSFs in a way that avoids including light from the companion.
For the region less than 1.5$\lambda$/D from the companion position, 
we replace the pixels in this region with the median value obtained for
all the pixels outside this region but at the same separation from the
central star (henceforth, the masked cube).  We then create PSFs to be subtracted from this 
masked cube by applying the PCA algorithm.  These PSFs are then
subtracted from the original cube, thus preventing self-subtraction of
the companion. The values obtained with this procedure and the value obtained from the 
unsubtracted datacube agree well. We evaluated the photometric error
by applying the same procedure in ten different points at the same separation from the central star and calculating the standard
deviation on these results. The same procedure was applied to the IRDIS data to obtain two more
spectral points.

\begin{figure}
\begin{center}
\centering
\includegraphics[width=8.0cm]{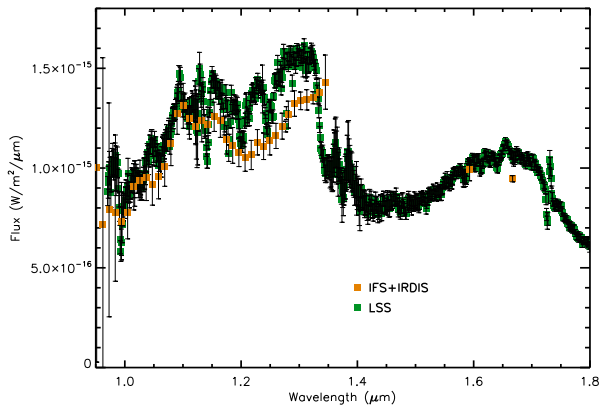}
\caption{{\bf Comparison between the spectrum of HR3549\,B obtained using IFS+IRDIS (orange squares) and the
one obtained using LSS (green squares).}}
\label{f:confifslss}
\end{center}
\end{figure}

We converted our spectrum from contrast into flux by multiplying it by
a flux calibrated BT-NEXTGEN \citep{2012RSPTA.370.2765A} synthetic
spectrum for the host star, adopting 
$T_{eff}$ = 10200\,K $\log{g}$=4.0 and [M/H]=0.0. We justify this 
choice of $T_{eff}$ and metallicity in Section~\ref{s:star}.
Finally, we smooth this spectrum to R=50 to match the resolution of 
the IFS spectrum. Given that the resulting spectrum has poorer
  S/N and resolution than the LSS IRDIS spectrum presented in
  Section~\ref{s:lss}, we do not perform fits to it with template spectra or synthetic spectra. However, 
this spectrum matches the LSS spectrum quite well {\bf as it it showed in Figure~\ref{f:confifslss}}.

\subsection{LSS}
\label{s:lss}

We applied the same procedure described for the IFS+IRDIS spectrum in
the previous section to the IRDIS LSS spectrum (resolution of R=350).
The spectrum was very noisy both at the short and long wavelength
extremes, thus we used only only the range between 0.97 and 1.8$\mu$m, 
for 729 measurements at distinct wavelengths as opposed to the 780
original spectral points.
We fit our final spectrum with spectra from both the Montreal Brown Dwarf and Exoplanet 
Library\footnote{\begin{tiny}\url{https://jgagneastro.wordpress.com/the-montreal-spectral-library/}\end{tiny}}
and the library from \citet{2013ApJ...772...79A}.  We convolved each library spectrum to match that of our
observed spectrum and interpolated to obtain flux values at the same
wavelengths as are covered by our {\bf LSS} spectrum.  The spectral
region between 1.35 and 1.45 $\mu$m is contaminated by a strong
telluric absorption band; this spectral region was hence left out of
our fit.    

\begin{figure*}
\begin{center}
\centering
\includegraphics[width=0.9\textwidth]{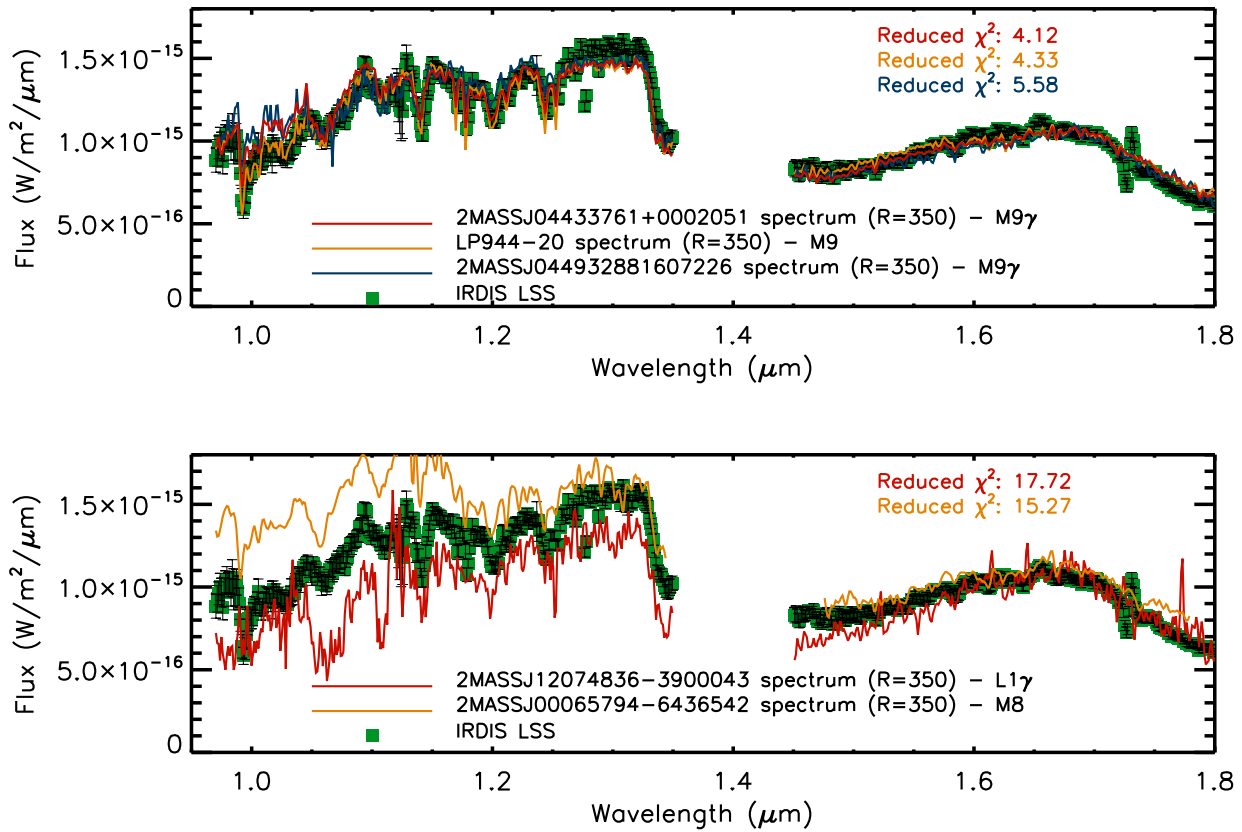}
\caption{The LSS medium resolution (R=350) spectrum is displayed with
  green squares, with error bars shown for each spectrum value.  
The three best-fit spectra from the spectral libraries described in
the text are also plotted.  {\it Lower panel:} Similar to the upper
panel, but with M8 and L1 spectra superimposed to illustrate the 
imperfect fit obtained in these cases.}
\label{f:fitlss}
\end{center} 
\end{figure*}

In Figure~\ref{f:fitlss} we display the medium (R=350) resolution spectrum obtained 
from the IRDIS LSS data. In the upper panel we display the three
best-fit spectra. The best fit
is obtained for a M9$\gamma$
\citep[2MASSJ02103857-3015313][]{2015ApJS..219...33G} and comparably
good fits are obtained for two other M9 objects, specifically
LP 944-20 \citep{2013ApJ...772...79A} and
2MASSJ044932881607226 \citep{2015ApJS..219...33G}. In the lower panel of Figure~\ref{f:fitlss}
we display the fit of our spectrum alongside two spectra of different spectral type. In this case we used
the L1 type object 2MASSJ12074836-3900043 \citep{2015ApJS..219...33G} and the M8 type object 
2MASSJ00065794-6436542 \citep{2015ApJS..219...33G}.


\section{Discussion}
\label{s:dis}

\subsection{Characteristics of HR\,3549}
\label{s:star}

According to the tables in \citet{2013ApJS..208....9P}, the colors of
the host star agree well with its classification as an AO star.  More
precisely, if E(B-V)=0.00 is adopted, a spectral type of
between a B9.5 and a A0 star is derived, but if E(B-V)=0.01 is adopted
instead, the colors are closer to those of a B9.5 spectral type.
This suggests that the reddening for this star is small, which is further confirmed by the
polarization measures of $0.030\%\pm0.030\%$
\citep{2000AJ....119..923H} and of $0.044\%\pm0.022\%$
\citep{2011ApJ...728..104S} for this star.
While a unique relationship between polarization and reddening does
not exist, we adopt a most probable relationship of
$E(B-V)\sim$$P_{int}/5$ from \citet{1975ApJ...196..261S}.
This leads to $E(B-V)=0.006\pm0.006$ 
from \citet{2000AJ....119..923H} and $E(B-V)=0.009\pm0.005$ from
\citet{2011ApJ...728..104S}. 
We thus adopt a best value for the reddening between 0.005 and 0.010
with an upper limit of 0.02. The impact of reddening on the spectral fit obtained in the previous Section
is thus negligible. \par
\subsubsection{Age of the HR\,3549}
\label{s:agestar}
A reliable determination of the age of the system is crucial for a proper characterization of the low mass component.
However, as pointed out by \citet{Mawet2015}, the star is not associated with any known young moving group. Moreover,
given its early spectral type, methods based on the activity, rotation and lithium abundance cannot be used to derive the age.
We searched several catalogs (Tycho2, UCAC4, PPMXL, SPM4.0) for wide common proper motion companions within 30 arcmin
from the star but did not identify any convincing candidates. 
Our kinematic analysis confirms the results by \citet{Mawet2015}; we also obtained space velocities 
U, V and W respectively of -16.7, -25.5 and -0.6 km/s.
Therefore the space velocities of HR\,3549 are well within the kinematic regions populated by young stars 
defined by \citet{2001MNRAS.328...45M} and very close to the kinematic boundaries of the Nearby Young Population
defined by \citet{2004ARA&A..42..685Z}. However, this result is not
conclusive as several old stars also share
these kinematic properties. \par
Therefore, we relied on isochrones fitting to derive the stellar age. We used the PARSEC models by \citet{2012MNRAS.427..127B} 
and the PARAM interface\footnote{\url{http://stev.oapd.inaf.it/cgi-bin/param}} (version 1.3) \citep{2006A&A...458..609D}.
This code uses as input the observational parameters
(effective temperature, trigonometric parallax, apparent magnitude in V band, and
metallicity along with their errors) to perform a Bayesian determination of the most likely
stellar intrinsic properties, appropriately weighting all the isochrone sections
which are compatible with the measured parameters.
A flat distribution of ages was adopted as a prior for this analysis.
The main stellar parameters obtained from our fits
are listed in Table~\ref{t:starpar}. The age reported in Table~\ref{t:starpar} obviously depends on the 
adopted metallicity. We assume [M/H]=0.00$\pm$0.10 as a reasonable estimate, since several studies have shown that the
metallicity of young stars in the solar neighborhood is consistent
with the solar value \citep[see e.g.][]{2006A&A...446..971J,2008A&A...480..889S,
2009A&A...501..553D,2011A&A...526A.103D,2011A&A...530A..19B,2012MNRAS.427.2905B}.\par
Our fit yields a most probable age of about 100-150 Myr, well within the range estimated by \citet{Mawet2015}.
However, the oldest possible values from \citet{Mawet2015} (300-400
Myr) are rejected by our fit. \par
\subsubsection{Rotation of HR\,3549}
\label{s:starrotation}
\begin{figure*}
\begin{center}
\centering
\includegraphics[width=0.9\textwidth]{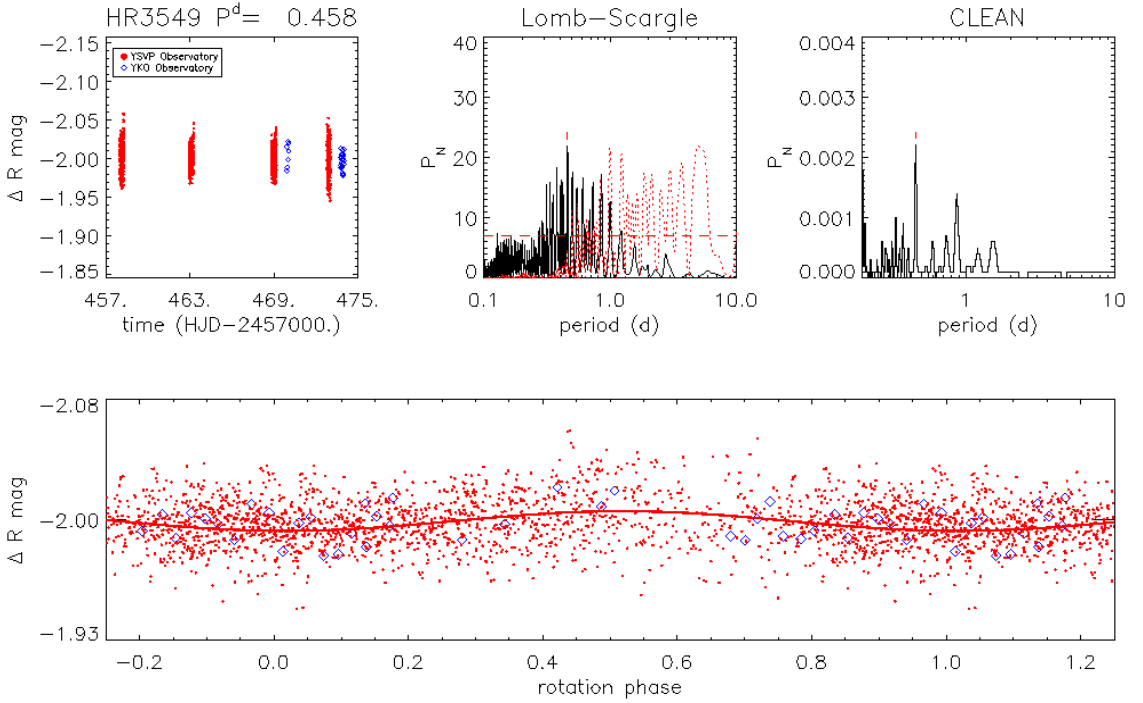}
\caption{Top left panel: differential R-band magnitude time series of HR\,3549. Top middle panel: LS periodogram. 
The dotted line represents the spectral window function. The horizontal dashed line is the power level corresponding 
to a 99.9\% confidence level. Top right panel: CLEAN periodogram. Bottom panel: light curve phased with the P = 0.458d 
rotation period. The solid line is a sinusoidal fit with amplitude $\Delta$R = 0.008 mag.}
\label{f:rotation}
\end{center} 
\end{figure*}

The non-SPHERE observations described in Section~\ref{s:nosphere} were then used to define the rotation of HR\,3549
using the following procedure. 
The  time series of differential magnitudes was analyzed using the Lomb-Scargle \citep[LS;][]{1982ApJ...263..835S} and 
the CLEAN \citep{1987AJ.....93..968R} periodograms to search for possible periodicities. We found the same most significant 
power peak (significance level $>$99.9\%)  at P = 0.458$\pm$0.005\,d
in both  LS  and CLEAN periodograms, with a light curve amplitude of $\Delta$R = 0.008\,mag. We consider this period as the stellar rotation period 
and attribute the slight rotational modulation to the presence of surface temperature inhomogeneities whose 
visibility is modulated by the stellar rotation. 
These results are displayed in Figure~\ref{f:rotation}. 
These results are unsurprising, as there is significant evidence of the existence of a large fraction ($\sim$40\%) 
of rotational variables among A type stars with light curve amplitudes
up to a few hundredths of a magnitude \citep{2016MNRAS.457.3724B}.
When the stellar rotation period measured above is combined with the stellar radius R = 1.88\,R$_\odot$ and the projected rotational velocity $v\sin{i}$ = 236$\pm$12\,km\,s$^{-1}$ \citep{2002A&A...393..897R}, we infer $\sin{i}$ = 1.13$\pm$0.1. Considering that HR\,3549 is expected to host some level of surface magnetic activity, and the $v\sin{i}$ measurement is not corrected for the additional broadening effects of macro- and micro-turbulence, the estimated  projected rotational velocity is likely an upper limit.
Therefore, we can assume that HR\,3549 is seen almost edge-on (i.e., $i$ $\sim$ 90$^{\circ}$).

\subsubsection{Separation of the disk}
\label{s:disksep}
{\bf The data from the IR excess allow determination of an approximate lower limit on the separation} of the disk from the star 
Indeed, having an excess at 22 \mic and no excess at 12 \mic we can consider the peak of the emission from
the disk around 22 \mic. Assuming a conservative error of 5 \mic on the position of the peak we can obtain, 
using the Wien law, an {\bf approximate} value for equilibrium temperature of the disk of $T_{eq}=131.7K$. From this value and assuming 
$T_{*}=10200$K and $R_{*}=1.88$\RSun we can calculate an {\bf approximate} radius of the disk of $R_{disk}=22.3$. 
We can then assume a value around 20 AU for the lower limit of the disk radius. \par 

\begin{table*}[!htp]
\caption{Stellar parameters obtained for HR\,3549 from the isochrones fitting assuming two different
values for E(B-V).}
\label{t:starpar}
\centering
\begin{tabular}{c c c c c c}
\hline
\hline
 E(B-V) & $T_{\rm eff}$  & Age (Gyr)  &  M/$M_{\odot}$ & $\log{g}$ & R/$R_{\odot}$  \\
\hline   
  0.00   &  $10079\pm200$    & $0.152\pm0.092$  & $2.326\pm0.071$  &  $4.228\pm0.032$ & $1.881\pm0.064$   \\
  0.01   &  $10314\pm200$    & $0.121\pm0.079$  & $2.375\pm0.070$  &  $4.243\pm0.031$ & $1.868\pm0.062$   \\
\hline
\end{tabular}
\end{table*}

\subsection{Characterization of HR\,3549\,B}
\label{s:companion}

Using the photometry defined in Table~\ref{t:photo} and exploiting the age range for the host star defined 
in Table~\ref{t:starpar}, we were able to estimate the companion mass using the BT-Settl evolutionary model \citep{2014IAUS..299..271A}. 
For our analysis we adopted five different ages, specifically 50, 100,
150, 200 and 300 Myr.  While our age analysis finds a most probable
age range of 100-150 Myr, younger and older ages cannot be completely
excluded.  Thus, we estimate mass for a broader range of ages here.
The age of 300 Myr is marginal but has been included in this analysis for completeness.   
We estimate mass separately for all 4 spectral channels covered (Y and J band from IFS and 
H2 and H3 band from IRDIS, mass estimates presented given in Table~\ref{t:mass}).
The mass determinations in different spectral channels agree well with
each other.   The companion mass ranges between $\sim$30~\MJup in the case of
a system age of 50 Myr and $\sim$70~\MJup in the case of a system age
of 300 Myr.   However, as discussed above,
we adopt a most probable age for the system of 100-150 Myr and thus a 
most probable mass for the companion of 40-50 \MJup.

\begin{table}
\caption{Mass determinations for HR\,3549\,B.  All masses are expressed in \MJup.}
\label{t:mass}
\centering
\begin{tabular}{c c c c c}
\hline
\hline
Age (Myr) & Y  & J  &  H2 & H3 \\
\hline   
   50   &  29    &  29    &   32  &  28    \\
  100   &  41    &  42    &   45  &  41    \\
  150   &  49    &  50    &   55  &  50    \\
  200   &  56    &  58    &   62  &  58    \\
  300   &  72    &  72    &   74  &  72    \\
\hline
\end{tabular}
\end{table} 

We estimate $T_{\rm eff}$ in a similar manner, with results presented in Table~\ref{t:teff}.
Adopting the same range of ages as before, $T_{\rm eff}$ varies between $\sim$2180\,K to $\sim$2500\,K. 
The values obtained for different spectral bands are again in good
agreement each other. For our most probable system age of 100-150 Myr, $T_{\rm eff}$ 
is between $\sim$2300\,K and $\sim$2400\,K. 

\begin{table}
\caption{$T_{\rm eff}$ determinations for HR\,3549\,B. All temperatures are in K.}
\label{t:teff}
\centering
\begin{tabular}{c c c c c}
\hline
\hline
Age (Myr) & Y  & J  &  H2 & H3 \\
\hline   
   50   &  2184    &  2181     &  2250  &  2134    \\
  100   &  2247    &  2263     &  2351  &  2249    \\
  150   &  2285    &  2310     &  2405  &  2307    \\
  200   &  2338    &  2365     &  2457  &  2364    \\
  300   &  2475    &  2485     &  2519  &  2488    \\
\hline
\end{tabular}
\end{table} 

\begin{figure}
\begin{center}
\centering
\includegraphics[width=8.0cm]{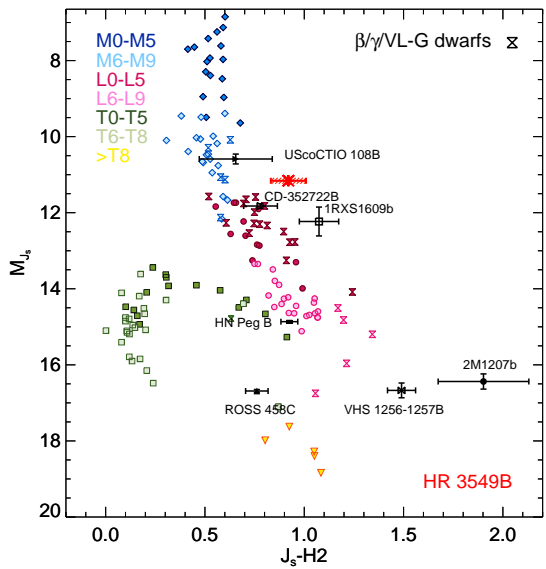}
\caption{Color-magnitude diagram showing the position 
HR\,3549\,B relative to that of M, L and T field dwarfs and of young
known companions.  HR\,3549\,B is indicated by the red
point.}
\label{f:colordiag}
\end{center} 
\end{figure}

In Figure~\ref{f:colordiag} we compare the position of HR\,3549\,B on a color-magnitude diagram with the positions
of M, L and T field dwarfs and young companions. The color-magnitude diagrams was generated using the synthetic SPHERE photometry of low gravity ($\beta$/$\gamma$/VL-G) dwarfs and old field MLTY dwarfs. This photometry was generated from the flux-calibrated near-infrared spectra of the sources gathered from the SpeXPrism library \citep{2014ASInC..11....7B}, and additional studies \citep{2010A&A...517A..76P, 2013ApJ...772...79A, 2014A&A...562A.127B, 2010ApJ...725.1405B, 2011ApJ...729..139W, 2015ApJ...804...96G, 2015ApJ...804...92S, 2014AJ....147...34S, 2015ApJ...799..203G, 2013ApJS..205....6M, 2013ApJ...777L..20L, 2010ApJ...719..497L, 2008A&A...482..961D}. \par
The position of
HR\,3549\,B is indicated by a red point and lies at the 
transition between M and L dwarfs, thus confirming the spectral classification M9-L0 obtained in 
Section~\ref{s:res}. However, the color of our object is redder than the field objects. This is in agreement with our estimation
of an young age for this system \citep{2008ApJ...689.1295K,2009AJ....137.3345C,2010AJ....139.1808S}.

\subsection{Fitting with synthetic spectra}
\label{s:synt}

\begin{figure*}
\begin{center}
\centering
\includegraphics[width=0.9\textwidth]{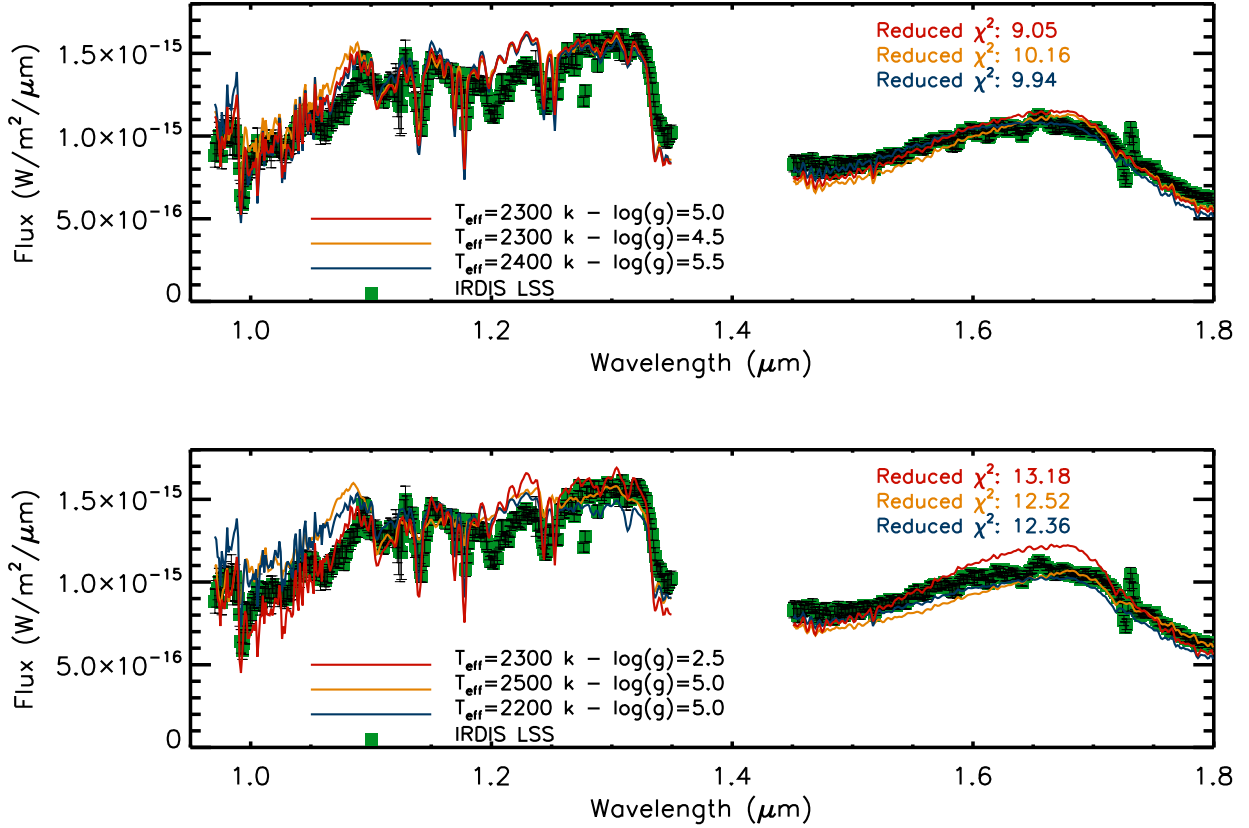}
\caption{Three best fitting synthetic spectra superimposed with the HR\,3549\,B 
LSS medium resolution spectrum. {\it Lower panel} Same as the upper panel but including synthetic spectra with worse fits.}
\label{f:syntlss}
\end{center} 
\end{figure*}

To further constrain the BD physical characteristics we fit its LSS spectrum 
with BT-Settl synthetic spectra models \citep{2014IAUS..299..271A}. 
As with the template spectra, we choose to fit only the LSS
medium resolution spectrum as it has significantly better S/N and resolution than the IFS+IRDIS spectrum. 
We again stress that the results obtained with the two different
spectra are in excellent agreement.
We have excluded the region between 1.35 and 1.45 $\mu$m from
the fit due to the strong telluric water absorption feature at these wavelengths.
In the upper panel of Figure~\ref{f:syntlss} we display the three best
fit synthetic spectra.  The best fit in this case is obtained with
a model with $T_{\rm eff}$=2300\,K and $\log{g}$=5.0. 
Models with $T_{\rm eff}$=2400\,K and $\log{g}$=5.5 and 
$T_{\rm eff}$=2300\,K and $\log{g}$=4.5 produce comparably good fits
to the spectrum. 
Thus,  we adopt $T_{\rm eff}$=2350$\pm$100\,K and 
$\log{g}$=5.0$\pm$0.5 for the companion.   
We integrated this model over all the wavelengths to calculate the
total flux of our object, in order
to obtain an estimate of the radius of the companion from 
Stefan-Boltzmann law. In this manner, we obtained a radius 
value of 1.13$\pm$0.17 \RJup where the uncertainty is driven 
primarily by uncertainty on the $T_{\rm eff}$, while the error on the
parallax is less important. 
{\bf The gravity and radius implied by our synthetic model fits exclude the youngest part of the 50-300 Myr age range considered above
given that they mean that the companion is quite evolved.} 
The value of $\log{g}$ in fact agrees well with what we would expect for an age of 100-150 Myr.
In this age range, we estimate a mass range of
40-50 \MJup (see Table~\ref{t:mass}), corresponding to 
$\log{g}$=4.89 and $\log{g}$=4.99 respectively. 
The temperature range of 2300-2400\,K obtained from the model fit is
also in good agreement with what we obtained for a 100-150 Myr object, as one can see from Table~\ref{t:teff}.


\subsection{Orbit determination for HR\,3549\,B}
\label{s:orbit}

We used the Least-Squares Monte-Carlo (LSMC) approach described in
\cite{Ginski2013} to determine if we could constrain the orbit of the
brown dwarf companion with the existing astrometry.
To reduce the wide parameter space, we set the system mass to a fixed value of 2.35\,M$_\odot$. This includes 2.3\,M$_\odot$ for the primary star (\citealt{2015ApJ...811..103M}) and an average value of 0.05\,M$_\odot$ for the companion. 
Note that the resulting orbital distributions are not highly sensitive to small changes in the companion mass. We further limited the semi-major axis to values smaller than 25.4\,arcsec (2350\,AU, 74.2 kyr period), following the criterion given in \cite{Close2003} for stability of the system against disruption in the galactic disk.\par
\begin{center}
\begin{figure*}[!htp]
\includegraphics[width=0.9\textwidth]{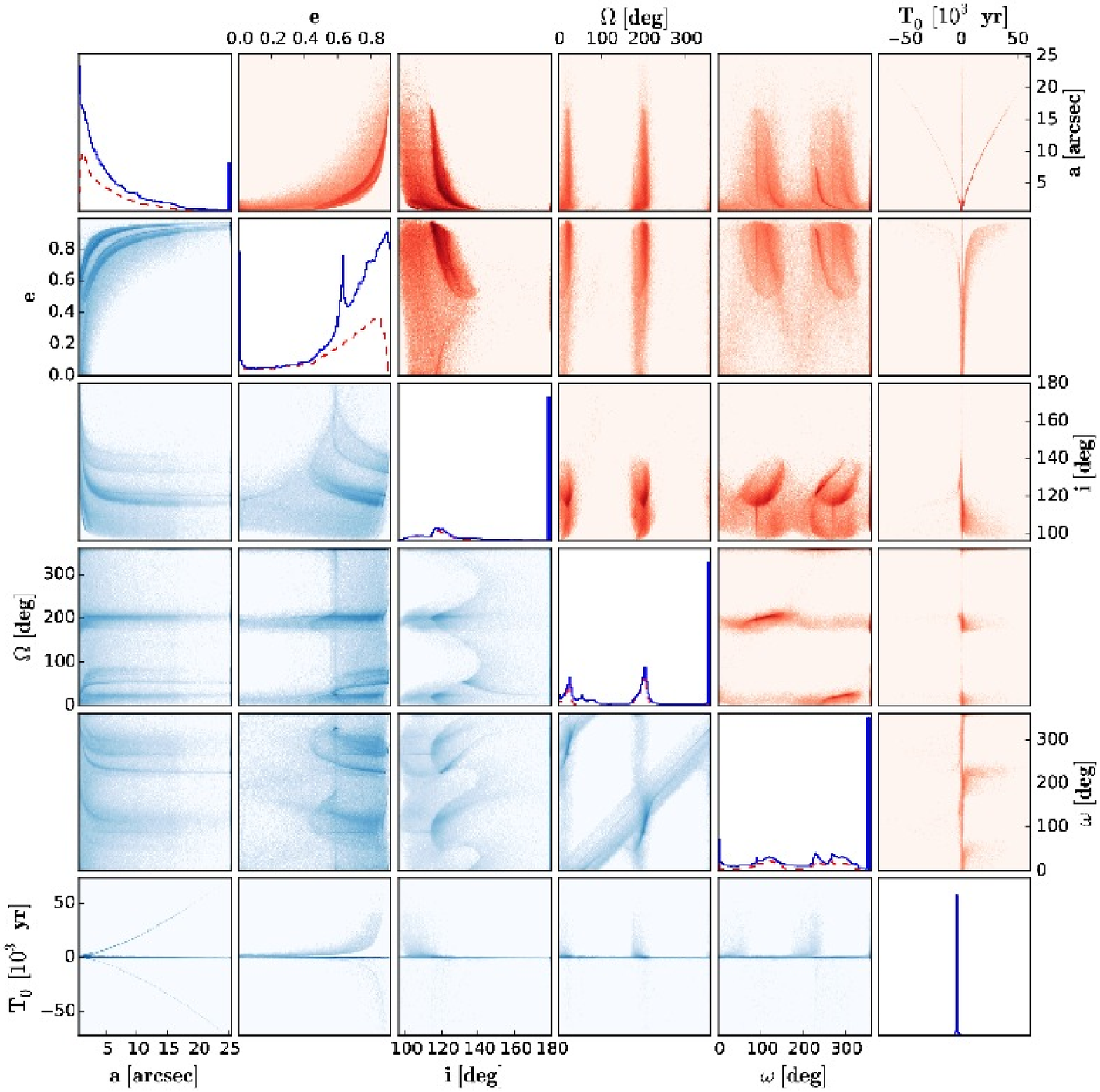}
\caption[]{Result of our LSMC orbit evaluation. On the diagonal the distribution for each individual orbit parameter is shown, while the other tiles show correlations between two orbit parameters. On the lower left in a blue hue we show all orbit solutions with a reduced $chi^2$ smaller than 2, out of 5$\times$10$^6$ orbit fits. 
On the upper right in a red hue we show all the solutions that are compatible with the presumed stable circumstellar disk.} 
\label{fig: results-lsmc}
\end{figure*}
\end{center}
\begin{center}
\begin{figure*}[!htp]
\includegraphics[width=0.45\textwidth]{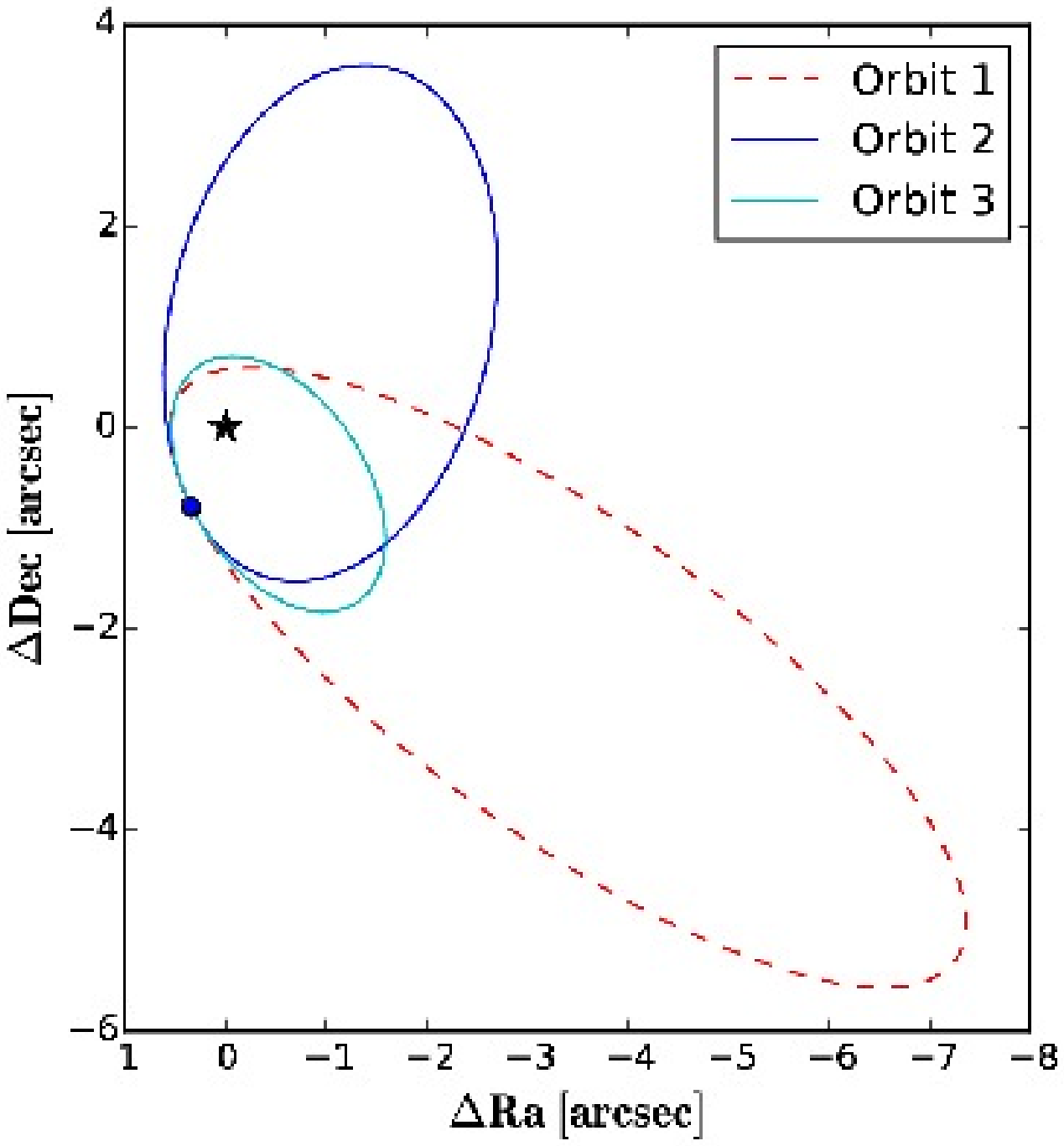}
\includegraphics[width=0.45\textwidth]{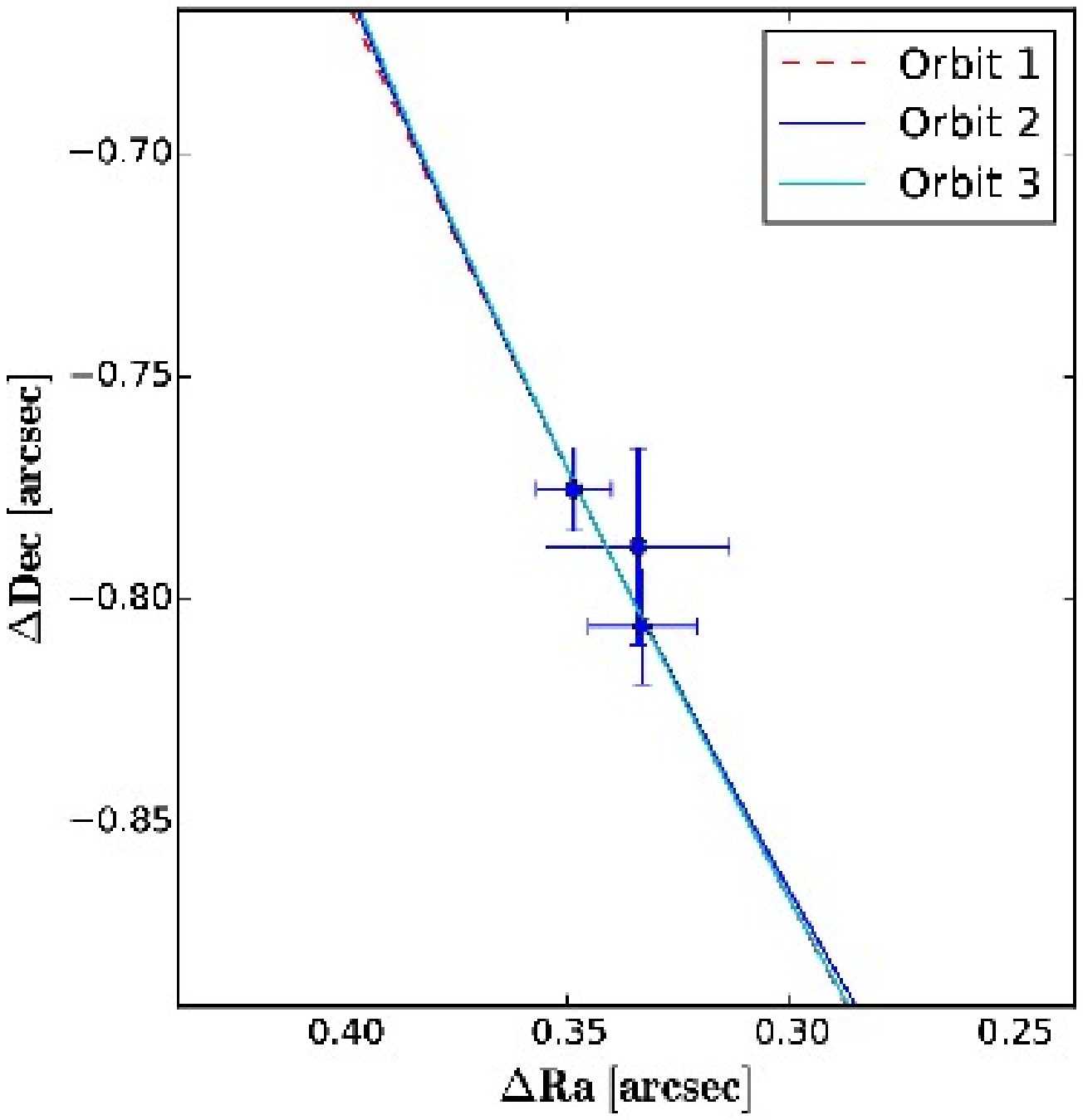}
\caption[]{{\it Left:} Top 3 best-fitting orbits out of 5$\times$10$^6$ runs of our LSMC fit for a mass of 1.22\,M$_\odot$. 
Solid lines represent the apparent orbits. Here the red dashed lines represents best fitting orbits that are not 
compatible with the presence of the disk while the blue lines represents orbit compatible with the disk. 
{\it Right: } Same image zoomed in on the data points. The corresponding orbit 
elements are listed in Table~\ref{tab: orbit-elements}.} 
\label{orbits} 
\end{figure*}
\end{center}

We then created 5$\times$10$^6$ random sets of orbital elements. Samples were drawn from uniform distributions of each orbital parameter. Each of these sets of orbital elements was then used as starting point of a Least-Squares minimization routine using the Levenberg-Marquardt algorithm.
We show the results of these 5$\times$10$^6$ fitting runs in the lower left of Fig.~\ref{fig: results-lsmc}. We include all solutions with a reduced $\chi^2$ smaller than 2. This rather large range was choosen to allow for potential systematic offsets between the astrometric measurements (i.e. due to different calibrators). 
This criterion is fulfilled by 433198 individual solutions. The three best fitting orbit solutions that we recovered are shown in Fig.~\ref{orbits}, and their orbital parameters are listed in Tab.~\ref{tab: orbit-elements}.\par
We find a large number of possible bound orbits that are compatible with the astrometric measurements. Specifically we find circular as well as eccentric orbits with eccentricities up to 0.993. 
It should be noted that it is of course quite common to find a large number of eccentric long period orbits that fit short orbital arcs without significant curvature. \par
In addition to the astrometric measurements, we consider the presence of an infrared excess of the host star to constrain the orbit of the companion. 
As seen in Section~\ref{s:disksep}, the infrared excess of the star is best fit by the presence of a circumstellar debris disk with a lower limit for the radius up to $\sim$20\,AU.
We use the formulas given by \citet{1999AJ....117..621H} to compute the critical semi-major axis for dynamical stability of
additional objects in the system given the eccentricity of an orbit solution below which the companion would disrupt the disk. 
We have to stress that the mass ratio of HR\,3549 is slightly outside the range over which the \citet{1999AJ....117..621H}
relationship were defined. However, detailed investigation of the dynamical stability of additional bodies for the three best
fit orbits listed in Table~\ref{tab: orbit-elements} was performed using the frequency map analysis (FMA) technique as in
\citet{2005LPI....36.1289M}, resulting in stability limits about 30\% larger than the \citet{1999AJ....117..621H} equation. 
We then exclude all the orbits that do not fulfill this conditions.
The result is shown in the upper right of Fig.~\ref{fig: results-lsmc}. Our best fitting orbit is indeed not compatible with the presence of this inferred disk as is indicated in Fig.~\ref{orbits}.
Since we only infer the presence of the circumstellar disk and its outer radius from unresolved photometry, we first discuss the orbital parameter distribution of the companion without the constraints introduced by the disk.

\subsubsection{Orbit without disk constraints}
\label{nodisk}

If the companion has formed in-situ, either via gravitational
instability in the protoplanetary disk or star-like via collapse in
the protostellar cloud, we would in principle expect a low orbital
eccentricity.  We indeed recover a large number of circular
orbits. The inclination of circular orbits can already be constrained
between 101.4\,deg and 137.5\,deg and orbital periods between 509\,yr
and 2579\,yr.  In addition, the longitude of the ascending node
$\Omega$ can be constrained to two peak values close to 0 or 180\,deg
for the circular case.\par
On the other hand eccentric orbits might be a sign that the companion
formed in a different part of the system and later experienced dynamic
interaction with either an additional companion or a close encounter
with another stellar object.  Of course high eccentricities could also
be explained in other ways, e.g. the companion could have formed
outside of the system and was later captured.  We find a large number
of such eccentric orbits. In fact all our best fitting orbits shown in
Fig.~\ref{orbits} are highly eccentric. In general eccentric solutions
fit the current astrometric data set better than circular orbits. If
we limit the sample of relevant orbits to fits with a reduced $\chi^2$
smaller than 1, then we find a minimum eccentricity of $\sim$0.3.\par
With increasing eccentricity the semi-major axis typically increases
as well to fit the astrometric data points, up to our {\bf upper limit for the semi-major axis}
of 25.4\,arcsec. However, there is also a large number of highly
eccentric orbit solutions with short orbital periods that we can not
yet exclude.  We observe a peak of the eccentricity at a value of
$\sim$0.65. Indeed the majority of orbit solutions at this
eccentricity peak have small semi-major axes of $\sim$0.9\,arcsec
(83.2\,AU, 495 yr period), but orbits with semi-major axes up to 13\,arcsec,
can not be excluded.  The range of possible inclinations for orbits at
this eccentricity peak goes from 96.7\,deg to 180\,deg, i.e. is
significantly wider than for circular orbits. The longitude of the
ascending node becomes essentially unconstrained at this point as
well. \par
To distinguish between potential formation scenarios for the brown
dwarf companion, it is important to check if 
the inclinations of its potential orbits are compatible
with the inclination of the stellar spin axis. One would expect that
an object that formed in-situ in the protoplanetary disk would show a
similar inclination of its orbit as the stellar spin axis, while
this is not necessarily true for an object that formed in a star-like
fashion.  Considering our conclusions about the value of
  $\sin{i}$ in Section~\ref{s:star}, we would expect highly inclined
orbits for the companion if its orbit is indeed aligned with the
stellar spin. As discussed earlier and also as noted in
Fig.~\ref{fig: results-lsmc}, we indeed recover a range of such highly
inclined orbits (circular and eccentric).  We can thus conclude that
the current astrometry is consistent with in-situ formation scenarios
of the companion in the protoplanetary disk. However, we also find a
variety of orbit solutions that are also consistent with other
formation scenarios. \par
Continued astrometric monitoring of this object over the next decade
might shed some light on its formation history, especially if
significant orbit curvature is detected. However, it will be very
  difficult if impossible to disentangle different potential orbits via
  observation in the next few years. For example, our current estimates 
suggest that we would have to wait until 2060 to fully distinguish between the
  three best fit orbits listed in Table~\ref{tab: orbit-elements}. On the other hand, new observations
  could help in narrowing down the parameter space for the companion.

\subsubsection{Orbit with disk constraints}
\label{disk}

If we consider the presence of the circumstellar debris disk (outer
radius of $\sim$20\,AU), we can further constrain the orbital
parameters of the companion.  In the presence of the disk, highly
eccentric orbits can not have arbitrarily small semi-major axes, since
the companion would then disrupt the disk at periastron passage.  One
immediate consequence is that the peak that we find at an
eccentricity of $\sim$0.65 vanishes completely. This is not surprising
given that most of the orbital solutions in this eccentricity range
had small semi-major axes of only $\sim$0.9\,arcsec (83.2\,AU, 495 yr
period).  The maximum eccentricity that we recover is 0.925 compared
to 0.993 for the unconstrained case. As is seen in Fig.~\ref{fig:
  results-lsmc} we lose the "steeper" branch of the semi-major axis -
eccentricity distribution beyond an eccentricity of $\sim$0.5.\par
The inclination can be constrained as well with 99.5\% of all
solutions now located between 96.4\,deg and 140\,deg. This is highly
consistent with the range of possible inclinations that we find
considering the possible inclinations of the stellar spin axis, and
might be an indication that the inclination of the companion orbit and
the stellar spin axis are indeed aligned.  Finally we can also
constrain the longitude of the ascending node to values of
192.2$\pm$7.2\,deg ($\pm$180\,deg depending on the radial velocity of the
companion).\par
Further astrometric monitoring of the system over the next decade will
significantly improve our understanding of the orbit of this companion and will thus
enable us to understand formation scenarios for this interesting
system.

\begin{table}
 \centering
  \caption{Orbit elements and $\chi^2_{red}$ of the best-fitting orbits shown in Fig.~\ref{orbits}.}
  \begin{tabular}{@{}lccc@{}}
  \hline   
	\hline
 	Nr. 			& 1			&  2 			& 3 		\\
 	\hline
	a\,[arcsec] 		& 4.77			& 3.24			& 1.44		\\
        a\,[AU]                 & 441.2                 & 299.7                 & 133.2         \\
	e			& 0.88			& 0.69			& 0.56		\\
	P\,[yr]			& 6034.6		& 3383.8		& 1002.3	\\	
	i\,[deg]		& 134.3			& 123.1			& 135.0		\\
	$\Omega$\,[deg]		& 53.9			& 182.9			& 26.1		\\
	$\omega$\,[deg]		& 0			& 63.9			& 336.8		\\
	T$_0$\,[yr]		& 2103.7		& 2042.5		& 2111.3	\\
	$\chi^2_{red}$		& 0.294			& 0.294			& 0.294		\\
 \hline\end{tabular}

\label{tab: orbit-elements}
\end{table}

\subsection{Mass limits for other objects in the system}
\label{s:masslimit}

\begin{figure}
\begin{center}
\centering
\includegraphics[width=8.0cm]{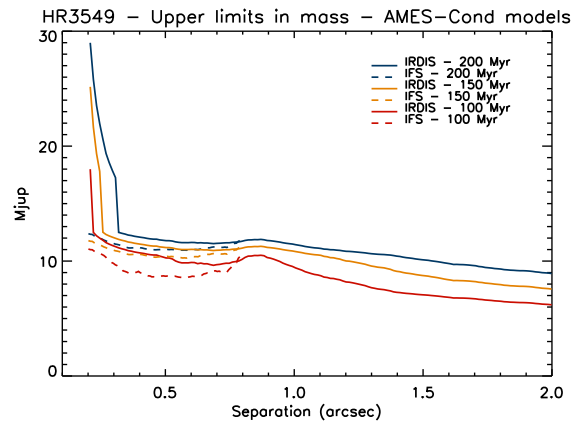}
\caption{Upper limits for possible additional components of the system versus separation from the central star.
Different ages are assumed: 200 Myr (blue line), 150 Myr (orange line) and 100 Myr (red line). We display both
upper limit obtained with IRDIS (solid lines) and IFS (dashed lines).}
\label{f:masslimit}
\end{center} 
\end{figure}

Using our derived contrast curve (Figure~\ref{f:contrast}), 
adopting our optimal age range of 100-150 Myr (see
Section~\ref{s:synt}), and converting from contrast to mass using the
Cond-Ames models, we set upper limits on the possible masses of
other components of the system.
The final result of this analysis is displayed in Figure~\ref{f:masslimit} where we show the mass limit obtained
both for IRDIS and IFS at three different possible system ages: 100,
150 and 200 Myr. While our previous analysis 
favours an age of 100-150 Myr,  we included the 200 Myr age here as a conservative case. For all the considered ages,
both IFS and IRDIS exclude the existence of any other sub-stellar objects with a mass larger than $\sim$13\MJup at separation 
less than 0.8 arcsec. For an age of 100 Myr, we can exclude the
presence of $>$9\MJup objects at 0.5 arcsec from our IFS observations.
From our IRDIS observations, we can exclude the presence of objects with mass larger than 10\MJup at separation greater than 1 arcsec.    

\section{Conclusions}
\label{conclusion}

We present here SPHERE observations of the star HR\,3549, recovering the low mass
companion discovered by \citet{Mawet2015} both with IFS and IRDIS. We obtained
precise astrometry and photometry in 4 bands for the companion.  Our
astrometry further confirms that the companion is bound to its host
star.   Significant uncertainty is still present in the age of this
system. Assuming a conservative age range of 50-300 Myr combined with
our multiband photometry,  
we estimate a mass range for this companion of  $\sim$30-70 \MJup and 
a $T_{\rm eff}$ range between $\sim$2200 and $\sim$2500\,K.
However, the position of this companion in the color-magnitude diagram as well
as isochrone fitting strongly suggests that this system is young
(100-150 Myr), thus, the most probable mass of the companion is between 40-50 \MJup while its
$T_{\rm eff}$ is probably between 2300-2400\,K. This latter values are further confirmed by fits with 
BT-Settl synthetic spectra. Our best fit BT-Settl model has $T_{\rm eff}$=2300\,K and 
$\log{g}$=5.0 but a comparably good fit is obtained with $T_{\rm
  eff}$=2400\,K and $\log{g}$=5.0. 
The BT-Settl fits agree well with the $T_{\rm eff}$ and mass determination obtained from the object photometry. \par
Fitting both the IFS+IRDIS and medium resolution IRDIS LSS spectra for
this object yielded a spectral type of M9-L0, right at the transition
between M and L dwarf. The M/L transition spectral type is further confirmed by the position of the companion on the 
color-magnitude diagram. \par
The astrometric position obtained from our analysis together with the two earlier positions from \citet{Mawet2015}
allowed us to simulate a suite of potential orbits and thus 
constrain the companion's orbital parameters. While the time span 
from these three astrometric points is too small to definitively
measure any orbital parameters, we were however
able to exclude families of orbits.  The current 
astrometry is consistent with in-situ formation scenarios for the companion within the protoplanetary disk.  However, 
we also find a variety of orbit solutions that are more consistent with other formation scenarios. Continued 
astrometric monitoring of this object over the next decade might shed
some light on its formation history, 
especially if significant orbit curvature is detected. SPHERE will be a
very valuable instrument for this continued astrometric monitoring. \par

Moreover, new data (for instance, the GAIA first data release) will
better define the age of the system, thus allowing us to 
place tighter constraints on the physical charateristics of HR\,3549\,B.


\begin{acknowledgements}
We are grateful to the SPHERE team and all the people at Paranal for the great effort during SPHERE GTO run. D.M., A.Z., A.-L.M., R.G., R.U.C., S.D. acknowledge support from the ``Progetti Premiali'' funding scheme of the Italian Ministry of Education, University, and Research. We acknowledge support from the 
French National Research Agency (ANR) through the GUEPARD project grant ANR10-BLANC0504-01. SPHERE is an instrument designed and built
by a consortium consisting of IPAG, MPIA, LAM, LESIA, Laboratoire Fizeau,
INAF, Observatoire de Gen\`eve, ETH, NOVA, ONERA, and ASTRON in collaboration with ESO.  \par
\end{acknowledgements}

\tiny This research has benefited from the Montreal Brown Dwarf and Exoplanet Spectral Library, maintained by Jonathan Gagn\'{e}.

\tiny {SPHERE is an instrument designed and built by a consortium consisting
of IPAG (Grenoble, France), MPIA (Heidelberg, Germany), LAM (Marseille,
France), LESIA (Paris, France), Laboratoire Lagrange (Nice, France), INAF-- Osservatorio di Padova (Italy), Observatoire de Gen\`eve (Switzerland), ETH
Zurich (Switzerland), NOVA (Netherlands), ONERA (France) and ASTRON
(Netherlands), in collaboration with ESO. SPHERE was funded by ESO, with
additional contributions from CNRS (France), MPIA (Germany), INAF (Italy),
FINES (Switzerland) and NOVA (Netherlands). SPHERE also received funding
from the European Commission Sixth and Seventh Framework Programmes as
part of the Optical Infrared Coordination Network for Astronomy (OPTICON)
under grant number RII3-Ct-2004-001566 for FP6 (2004-2008), grant number
226604 for FP7 (2009-2012) and grant number 312430 for FP7 (2013-2016).}

\bibliographystyle{aa}
\bibliography{hip43620_v3l}

\end{document}